\documentstyle[preprint,aps]{revtex}

\begin{document}
\title
{The Ising Transition in the double-frequency sine-Gordon model}     
\author{Fei Ye\hspace{1.0cm}Guo-Hui Ding\hspace{1.0cm}Bo-Wei Xu}
\address
{Department of Physics, Shanghai Jiao Tong University, Shanghai 200030,China} \date{\today}
\draft
\maketitle
\begin{abstract}
In the present paper we utilize the renormalization group(RG) technique to analyse the 
Ising critical behavior in the double frequency sine-Gordon model. The one-loop RG equations
obtained show unambiguously that there exist two Ising critical points besides the trivial
Gaussian fixed point. The topology of the RG flows is obtained as well.
\end{abstract}

\newpage
\narrowtext
The sine-Gordon(SG) model plays an important role in the condensed matter physics
since many one-dimensional physical system can be mapping onto it. The SG hamiltonian
can be written as a Gaussian model of a scalar field $\phi$ perturbed by a vertex
operator $\cos\beta\phi$, which has the form
\begin{equation}
{\cal H}_{SG}={\cal H}_{Gauss}+g'\cos\beta\phi\;\label{SG}
\end{equation}
\begin{equation}
{\cal H}_{Gauss}={1\over 2}\pi^2+{1\over 2}(\partial_x\phi)^2\;\label{Gauss}
\end{equation}
with coupling constant $g'$. As well known, for $\beta^2>8\pi$, the perturbation is 
irrelevant and the system is gapless, and for $0<\beta^2<8\pi$, the vertex operator
becomes relevant and drive the system to a strong-coupling massive phase with a 
mass gap $m\sim g^{4\pi/(8\pi-\beta^2)}$\cite{Tsvelik}.

Recently, the double-frequency sine-Gordon(DSG) model has received much attention
which is the SG model subjected to another vertex operator
\begin{equation}
{\cal H}_{DSG}={\cal H}_{Gauss}+g'\cos\beta\phi+\lambda'\cos{\beta\over 2}\phi\;\label{DSG}.
\end{equation}
This model was found to display an Ising criticality with central charge $c=1/2$ on a quantum
critical line $\lambda'=\lambda'_c(g)$(quasi-classically $\lambda'_c(g)=4g'$) by Delfino and 
Mussardo\cite{Delfino}. They also argued that this Ising transition is a universal property
of the DSG model\cite{Delfino} as long as $\beta^2<8\pi$. Fabrizio ${\it et.al}$ investigated
in detail the critical properties of this transition by mapping the DSG model onto the deformed
quantum Ashkin-Teller(DAT) model. Under a new representation of the DAT model, they succeeded
in identifying those degrees of freedom that become critical, and they also discussed the 
application of the Ising transition in some physical realization of the DSG model.

In this letter, we shall give the renormalization group(RG) analysis of the Ising criticality
in the DSG model. In the following, we assume that $32\pi/9<\beta^2<8\pi$, so that both the 
vertex operators in Eq.(\ref{DSG}) are relevant and no other relevant operators are generated 
upon renormalization. It is clear that both these two vertex operators will make the theory
fully massive, if acting alone. However if they coexist, it will become gapless on some critical
line. The RG equations obtained explore this feature in an explicit way
and give the RG flows in the coupling parameters plane, in which
there exist three fixed points  to the RG equations including
the trivial Gaussian fixed point and two non-trivial Ising fixed points. 

To construct the one loop RG equations, we employ the formalism in Refs.\cite{Cardy1,Cardy2}
which is based on the operator product expansion(OPE) in real space. 
Firstly, we replace the coupling constants $g'$ and $\lambda'$ by 
the dimensionless bare coupling constants $g=a^{\Delta_g-2}g'$ and $\lambda=a^{\Delta_\lambda
-2}\lambda'$ with $\Delta_g=4\Delta_\lambda=\beta^2/(4\pi)$ and $a$ being the microscopic
short distance cut-off. 
In our case, the coefficients of the following OPEs are needed to construct the RG 
equations
\begin{eqnarray}
:\cos{\beta\over 2}\phi(z,{\bar z})::\cos{\beta\over 2}\phi(w,{\bar w}):&\sim&
{1\over 2}|z-w|^{\beta^2\over 8\pi}:\cos\beta\phi(w,{\bar w}):-{\beta^2\over 8}|z-w|^{2-{\beta^2
\over 8\pi}}:\partial_w\phi\partial_{\bar w}\phi:+\cdots\;,\label{ope1}\\
:\cos\beta\phi(z,{\bar z})::\cos\beta\phi(w,{\bar w}):&\sim&
{1\over 2}|z-w|^{\beta^2\over 2\pi}:\cos 2\beta\phi(w,{\bar w}):-{\beta^2\over 2}
|z-w|^{2-{\beta^2
\over 2\pi}}:\partial_w\phi\partial_{\bar w}\phi:+\cdots\;,\label{ope3}\\
:\cos\beta\phi(z,{\bar z})::\cos{\beta\over 2}\phi(w,{\bar w}):&\sim&{1\over 2}
|z-w|^{-{\beta^2\over 4\pi}}:\cos{\beta\over 2}\phi(w,{\bar w}):+\cdots\;.\label{ope2}
\end{eqnarray}
It should be noted that in Eqs.(\ref{ope1},\ref{ope3},\ref{ope2}) we only pick up the operators 
which have contribution to the RG equation.

Through Eqs.(\ref{ope1},\ref{ope2}) one can obtain the one-loop RG equations\cite{Cardy1,Cardy2}
\begin{eqnarray}
&&{dg\over d\ln l}=(2-\Delta_g)g-{\pi\over 2}\lambda^2\label{RGE1}\\ 
&&{d\lambda\over d\ln l}=(2-\Delta_\lambda)\lambda-\pi g\lambda\label{RGE2}
\end{eqnarray}
under the scaling transformation $a\sim e^l a$. 
The second term in Eqs.(\ref{ope1},\ref{ope3}) leads to the renormalization of $\beta^2$ 
\begin{equation}
{d\beta^2\over d\ln l}=-{\pi\over 4}\beta^4(g^2+{\lambda^2\over 4})\;,\label{beta}
\end{equation}
which results in the modification of $\Delta_g$ and $\Delta_\lambda$ in Eqs.(
\ref{RGE1},\ref{RGE2}). However, this higher order correction can be neglected since it
will not change our conclusion qualitatively for small $g$ and $\lambda$.

The zeros of the RG equations (\ref{RGE1},\ref{RGE2}) give the fixed points of the system, 
which include two Ising critical points $(g_c,\pm\lambda_c)$ with 
\begin{eqnarray}
&&g_c={1\over\pi}(2-\Delta_\lambda)\;\\
&&\lambda_c={1\over\pi}\sqrt{2(2-\Delta_g)(2-\Delta_\lambda)}\;
\end{eqnarray}
and the trivial Gaussian fixed point $g=\lambda=0$. 
The topology of the RG flows implied by Eqs.(\ref{RGE1},\ref{RGE2}) is plotted in Fig.(1).

In the vicinity of the Gaussian fixed point which is unstable with respect to the parameters
$g$ and $\lambda$, the RG trajectories have the form 
$\lambda\sim g^\mu$ with $\mu=(2-\Delta_\lambda)/(2-\Delta_g)$ and flow outward. 

In order to extract the asymptotic behavior of the RG flows near the Ising critical
points, one can substitute $g=g_c+\tilde{g}$ and $\lambda=\lambda_c+\tilde{\lambda}$ into
the RG equations and neglect the quadratic terms of $\tilde{g}$ and $\tilde{\lambda}$,
which leads to
\begin{eqnarray}
&&{d\tilde{g}\over d\ln l}={\pi\lambda^2_c\over 2g_c}\tilde{g}-\pi\lambda_c\tilde{\lambda}\;,
\label{icp1}\\
&&{d\tilde{\lambda}\over d\ln l}=-\pi\lambda_c\tilde{g}\;.\label{icp2}
\end{eqnarray}
These two equations (\ref{icp1},\ref{icp2}) can be rewritten in a decoupled form with the 
new variables $\xi_\pm=\tilde{g}+a_\pm\tilde{\lambda}$ as
\begin{equation}
{d\xi_\pm\over d\ln l}=\pi\lambda_c a_\mp \xi_\pm\;,\label{ddd} 
\end{equation}
where $a_\pm=(\lambda_c/g_c\pm\sqrt{16+\lambda^2_c/g^2_c})/4$.

Clearly Eq.(\ref{ddd}) implies that $\xi_-$ is relevant and will opens a mass gap proportional
to $\xi_-^{1/\pi\lambda_c a_+}$ and $\xi_+$ is irrelevant. 
As seen from Fig.(1), the Ising fixed point $(g_c,\lambda_c)$ is unstable with respect to
$\xi_-$ and stable with respect to $\xi_+$. The lines $\xi_+=0$ and $\xi_-=0$ act as separatrix
dividing the region around the Ising fixed point into four parts. Fig.(1) also indicates
there exist a critical line connecting the Gaussian fixed point and the Ising fixed point, any
point on which will flow toward the Ising fixed point and the central charge of the system
changes from $c=1$ to $c=1/2$
according to the Zamolodchikov's C-theorem\cite{Cardy1,Cardy2,Zamolodchikov}. 
Thus, all the initial hamiltonian on this line are in the same university class.
This critical trajectory has the form $\lambda=\zeta g^\mu$ with some fixed parameter
$\zeta>0$ near the Gaussian fixed point and has a linear asymtotic form $\xi_-=0$, namely, 
\begin{equation}
g+a_-\lambda-g_c-a_+\lambda_c=0
\end {equation}
around the Ising fixed point $(g_c,\lambda_c)$.

As for the other Ising fixed point $(g_c,-\lambda_c)$, the similar analysis gives that this
fixed point is unstable with respect to $\xi_+$ and stable with respect to $\xi_-$.
The critical line connecting this Ising fixed point and the Gaussian fixed point also has the 
form $\lambda=\zeta g^\mu$ but the parameter $\zeta<0$ near the Gaussian fixed point, and in the vicinity of the 
fixed point $(g_c,-\lambda_c)$ it becomes $\xi_+=0$, namely
\begin{equation}
g+a_+\lambda-g_c+a_+\lambda_c=0\;.
\end{equation}

As a conclusion, we employ the RG approach to study the Ising criticality occurring in the 
DSG model which finds some interesting applications in the 1D strongly correlated electron
systems. Two Ising fixed points are found besides the trivial Gaussian fixed point. The quantum
critical line $\lambda=\lambda_c(g)$, on which the system displays Ising criticality with 
central charge $c=1/2$, is obtained near the fixed points together with the RG flows.

The project is supported by the National Natural Science Foundation of China(Grant No.19975031)
and the RFDP(Grant No.199024833).


\begin{references}
\bibitem{Tsvelik} A.O.Gogolin, A.A.Nersesyan and A.M.Tsvelik 1998 {\it Bosonization and 
Strongly Correlated Systems} (Cambridge University Press)
\bibitem{Delfino} G.Delfino and G.Mussardo 1998 {\it Nucl. Phys.} B {\bf 516} 675
\bibitem{Fabrizio} M.Fabrizio, A.O.Gogolin and A.A.Nersesyan 2000 {\it Nucl. Phys.} B {\bf 580} 647 
\bibitem{Cardy1} J.L.Cardy 1996 {\it Scaling and Renormalization in Statistical Physics} (Cambridge 
University Press) 
\bibitem{Cardy2} J.L.Cardy 1988 {\it Lectures delivered at Les Houches Summer School on  Fields, Strings and Statistical Mechanics}
\bibitem{Zamolodchikov} A.B.Zamolodchikov 1986 {\it JETP. Lett.} {\bf 43} 730
\end{references}
\end{document}